\documentclass[10pt,letterpaper,onecolumn]{article}

\usepackage[letterpaper,top=1in,bottom=1in,left=0.75in,right=0.75in]{geometry}

\usepackage{times}

\usepackage{graphicx}
\usepackage{url}
\usepackage{amsmath,amssymb,amsfonts,amsthm}
\usepackage{geometry}
\usepackage{graphicx}
\usepackage{fancyvrb}
\usepackage{booktabs}

\title{Quantifying Azure RBAC Wildcard Overreach}
\author{
	Christophe Parisel\thanks{Email: ch.parisel@gmail.com}
}

\begin{document}

\maketitle

\begin{abstract}
Azure’s Role-Based Access Control (RBAC) leverages wildcard permissions to simplify policy authoring, but this abstraction often obscures the actual set of allowed operations and undermines least-privilege guarantees. We introduce Belshazaar, a two-stage framework that targets both (1) the effective permission set problem—deriving the exact list of actions granted by arbitrary wildcard specifications—and (2) the evaluation of wildcards permissions "spread". First, we formalize Azure action syntax via a context-free grammar and implement a compiler that expands any wildcard into its explicit action set. Second, we define an ultrametric diameter metric to quantify semantic overreach in wildcard scenarios. Applied to Microsoft’s official catalog of 15,481 actions, Belshazaar reveals that about 50\% of actions admit a cross Resource Provider reach when associated with non obvious wildcards, and that effective permissions sets are effectively computable. These findings demonstrate that wildcard patterns can introduce substantial privilege bloat, and that our approach offers a scalable, semantics-driven path toward tighter, least-privilege RBAC policies in Azure environments.
\end{abstract}

\section{Introduction}
Microsoft Azure's Role-Based Access Control (RBAC) system governs access to cloud resources through permission policies (e.g. Microsoft.Compute/* or */read) that specify allowed and denied operations. To simplify authoring of Built-in roles, Microsoft uses wildcard patterns in both \texttt{Actions} and \texttt{NotActions} clauses, enabling administrators to grant broad categories of Control Plane permissions with concise expressions. A similar approach is followed in the Data Plane, featuring wildcards in both \texttt{dataActions} and \texttt{NotDataActions} clauses.

\medskip
\noindent

Wildcards usage should be restricted to a very limited number of users (break glass accounts) and service principals (CD/CI pipeline). The vast majority of principals should be granted explicit \texttt{Actions} because implicit permissions introduce significant security risks: a single asterisk (\texttt{*}) can inadvertently grant access to hundreds of operations, violating the principle of least privilege~\cite{saltzer1975protection}.

\subsection{Findings}
\begin{itemize}
	\item Our research unravels that, through the design of Custom Roles, customers are given more flexibility in wildcards positioning than what Microsoft itself uses for making Built-in roles: they are entitled to place wildcards at infix locations, yielding a wide-ranging, hard to predict, permission reach. 

	\item A review of all Azure \texttt{Actions} also shows that, when combined with ill-placed wildcards, 50\% have a potential to get an extreme (cross Resource Provider) reach.
\end{itemize}

\newpage

\section{Problem Statement}
\subsection{Wildcards Expansion In Azure}

Current Azure RBAC APIs and Portal tools provide limited visibility into wildcard expansion, leaving security teams unable to:
\begin{enumerate}
\item Precisely determine which atomic operations are granted by wildcard patterns
\item Quantify the security impact of over-approximated permissions
\item Detect when Microsoft's service evolution expands existing wildcards
\item Substitute wildcard expressions with explicit permissions while preserving intended functionality
\end{enumerate}

\subsection{Our Approach}

We address these challenges through a formal language-theoretic framework that treats Azure RBAC actions as a grammar generating nodes and leaves in a structured hierarchy tree. Azure's dot-and-slash-delimited naming convention populates a natural tree structure over which we can define precise distance ultrametrics. By reverse-engineering Azure RBAC's wildcards expansion logic, we can generate an arbitrary number of valid wildcard actions that we treat as leaves of this tree. We can enslave this generation process to a genetic algorithm to find the most wide ranging actions with regards their respective locations in the tree.

\medskip
\noindent

This paper makes the following contributions:
\begin{itemize}
\item \textbf{Azure RBAC Actions Grammar}: We propose a simple grammar specification of Azure's actions syntax
\item \textbf{Wildcard Risks Quantification}: We define a mathematically principled ultrametric diameter for quantifying permission over-approximation granted by wildcards
\item \textbf{Automated Identification of Extreme Actions}: We identify the most risky permissions by using a genetic algorithm trained to minimize the ultrametric diameter
\item \textbf{Open-Source Implementation}: We present Belshazaar, a wildcards expansions compiler based on the aactions grammar. 
\end{itemize}

\section{Background}

\subsection{Azure RBAC Architecture}

Azure RBAC operates on a hierarchical namespace where segments are delimited by slash. Each operation follows the pattern:
\texttt{ResourceProvider/OperationsLevel1/.../OperationsLevelN/ActionVerb}

\medskip
\noindent

The first and last segments have a special meaning: the first segment declares the resource provider, whereas the last segment holds the actual action verb (read, write, delete, action). All other segments define intermediate levels of nesting. All segments are optional

For example, \texttt{Microsoft.Compute/virtualMachines/start/action} represents the action to start a virtual machine within Microsoft's Compute resource provider. Wildcards can appear in any segment except in the last one unless they are the only character in the segment. Here are some example of valid wildcard placements:
\begin{itemize}
\item \texttt{Microsoft.Compute/*} - All Compute operations
\item \texttt{Microsoft.*/read} - All read operations across Microsoft Resource Providers
\item \texttt{*/delete} - All delete operations 
\item \texttt{*} - All operations 
\end{itemize}

\subsection{Related Work}

Access control analysis has been extensively studied in security literature. Sandhu et al.~\cite{sandhu1996role} formalized the RBAC model, while Li et al.~\cite{li2003critique} identified fundamental limitations in role-based systems.

\medskip
\noindent

Cloud-specific access control has received growing attention. AWS has developed Zelkova~\cite{Zelkova}, an internal service that uses automated reasoning to analyze IAM policies. This approach allows AWS to verify that policies meet security requirements and do not grant unintended access. They also open-sourced Cedar~\cite{Cedar}, which lets Cloud customers perform a similar analysis on their custom IAM policies. Recently, David Kerber released IAM Lens~\cite{Lens}, a tool for evaluating AWS IAM policies offline. While AWS tooling is well covered in the literature, Azure has received little dedicated treatment. One of our recent works~\cite{Blast} focused on assessing lateral motion in Azure Data Plane leveraging a similar ultrametric in Azure Tenant containers hierarchy.

\medskip
\noindent

Formal methods in security policy analysis have been explored by Fisler et al.~\cite{fisler2005verification}. Jackson~\cite{jackson2012software} introduced the notion of Machine Diameter. We build upon this foundation by applying language theory specifically to hierarchical permission systems.

\section{Azure Actions Grammar}
\label{sec:grammar}

Through analysis of what Azure's Custom Roles Definition permits and denies, we reverse-engineered the grammar governing action strings to build our own PLY\cite{PLY} wildcards compiler.

\subsection{Lexical Analysis}
We found that modeling Azure permissions with only 3 tokens is enough for our needs: WILDCARD, SLASH and TEXT.

\begin{itemize}
\item \textbf{SLASH}: single \texttt{/} character.
\item \textbf{WILDCARD}: single asterisk \texttt{*} character, can appear only once. 
\item \textbf{TEXT}: matches sequences of alphanumerial characters (including dot, dash, underscore, dollar, the curly brackets, excluding all other characters).
\end{itemize}

Here is the PLY tokens formulation:

\medskip
\noindent

\begin{Verbatim}[fontsize=\small]
tokens = ('TEXT', 'WILDCARD', 'SLASH')

t_WILDCARD = r'\*'
t_SLASH   = r'/'

def t_TEXT(t):
    r'[-a-zA-Z0-9._{}$]+'
    return t

t_ignore = ' \t\n'
\end{Verbatim}

\subsection{Syntactic Analysis}

\begin{itemize}
	\item \textbf{SLASH}: acts as segments separator. Can be placed anywhere\footnote{although two slashes in a row and two dots in a row are permitted by Azure, we do not integrate them into our model for simplicity.}, except as a trailing character. 
\item \textbf{WILDCARD}: can be placed anywhere, except in the last segment unless it is the only character in the last segment. Spans one or more segments.
\item \textbf{TEXT}: can be placed anywhere. The non-alphanumerical characters are not metacharacters, they can be treated like alphanumerical ones.
\end{itemize}

\subsubsection{Preprocessing rules}
To make parsing straightfoward, before running the compiler we ensure that:
\begin{enumerate}
	\item case is made insensitive
	\item duplicate inputs are removed
	\item no more than one wildcard shows up in the input string
	\item a wildcard doesn't show up in the last segment, except if is the only character
	\item the last segment must be one of read, action, write, delete or a wildcard
	\item we ignore input actions without any segments (eg: '*')
\end{enumerate}

\subsubsection{Syntax rules}

\begin{Verbatim}[fontsize=\small]
pattern: segment_list
segment_list : segment
segment_list : segment_list SLASH segment
segment : TEXT
        | WILDCARD
	| TEXT WILDCARD
	| WILDCARD TEXT
	| TEXT WILDCARD TEXT
\end{Verbatim}

\subsubsection{Grammar Properties}

This grammar exhibits several important properties:
\begin{itemize}
\item \textbf{Grammar Completeness}
Our specification accepts all valid Azure permission strings pulled from Azure Resource Provider API. These permissions are all explicit: they contain no wildcards. At the time of writing, we validated completeness by parsing all 15,481 control plane actions without syntax errors or illegal characters.

Our specification accepts wildcards from common built-in Owner and Contributor roles.

\item \textbf{Wildcard Expressiveness}
Through testing of wildcard patterns at various locations, we confirmed that Azure supports:
\begin{itemize}
	\item Prefix wildcards: {\small \texttt{Microsoft.Stor*} matches \texttt{Microsoft.Storage/locations/usages/read} }
	\item Suffix wildcards: {\small \texttt{*Machines/reimage/action} matches \texttt{Microsoft.Compute/virtualMachines/reimage/action} }
	\item Infix wildcards: {\small \texttt{Micr*ft.AAD/Operations/read} matches \texttt{Microsoft.AAD/operations/read} }
\item Complete wildcards: \texttt{*} matches any single segment
\end{itemize}
\end{itemize}

\subsubsection{Production rules}
The semantics rules of our compiler are available in appendix.

\section{Wildcards Analysis Framework}
\label{sec:framework}

We now present our framework for analyzing wildcards in Azure RBAC: we treat each wildcard expression as a regular language\cite{regular_languages} over the Azure actions alphabet. Using all permissions strings from the Azure Resource Provider API, we are able to infer this alphabet exhaustively. With the alphabet now formulated, we can generate explicitly all \texttt{Actions} for each of these languages. 

We do the same for \texttt{NotActions}. Substracting \texttt{NotActions} from an \texttt{Action} yields the effective permission set of this \texttt{Action}.

\subsection{Mathematical Foundations}

\subsubsection*{Azure Operation Universe}
Let $\mathcal{U}$ denote the universe of all Azure operations, where each operation $u \in \mathcal{U}$ follows the canonical form:
$$u = \text{segment}/.../\text{segment}/.../\text{actionVerb}$$

\subsubsection*{Wildcard Language}
Given a wildcard expression $w$, define language $L(w) \subseteq \mathcal{U}$ as the set of all operations matched by $w$. Formally:
$$L(w) = \{u \in \mathcal{U} : u \text{ matches glob pattern } w\}$$

If $u$ is an Azure permission without a wildcard, $L(u)=\{ u \}$.

\subsubsection*{Permission Set}
In Azure RBAC, permission sets are defined at the role definition level. A role definition encompasses a set of \texttt{Actions} and a set of \texttt{NotActions}. Since both sets are entirely independent, to determine the permission set of a single Azure action in the \texttt{Actions} set of a role definition, we need to substract all the \texttt{NotActions}.

For an \texttt{Action} $A$ and \texttt{NotActions} $N = \{n_1, n_2, \ldots, n_m\}$, the permission set is:
$$P(A, N) = A \setminus \left(\bigcup_{j=1}^{m} n_j\right)$$

Since P may contain wildcards in $A$ and $n_i$, P is not effective. We expand all wildcards treating each $A$ and $n_i$ as a sentence of Azure's permissions grammar.

The effective, computable permission set becomes:
$$P_{eff}(A, N) = L(A) \setminus \left(\bigcup_{j=1}^{m} L(n_j)\right)$$

\subsection{Ultrametric Distance}

To quantify permission over-approximation induced by wildcards, we leverage Azure's hierarchical namespace structure and define its ultrametric distance.

\subsubsection*{Hierarchy Tree}
Model Azure's permission namespace as a tree $T$ where:
\begin{itemize}
\item Root (level 0) represents the global namespace
\item Level 1 nodes represent Microsoft resource providers (e.g., \texttt{Microsoft})
\item Level 2 nodes represent the sub providers (e.g., \texttt{Storage})
\item Level n nodes represent resource types and operations groups (e.g., \texttt{storageAccounts})
\item Leaves represent action verbs (e.g., \texttt{read})
\item Levels are segmented using a separator\footnote{We use both the slash and the dot to delimit segments. While using dot is completely optional, it provides a finer granularity than using slashes alone.}
\end{itemize}

\subsubsection*{Ultrametric Distance}
For operations $u, v \in \mathcal{U}$, define:
$$d(u, v) = \text{depth}(\text{LCA}(u,v))$$

\medskip
\noindent

where $\text{LCA}(u,v)$ is the lowest common ancestor of $u$ and $v$ in tree $T$. Root has depth 0, Microsoft resource providers have depth 1, etc.

\medskip
\noindent

Readers familiar with\cite{Blast} will notice that unlike the exponantial Blast Radius ultrametric, our ultrametric is linear.

Here is a first example:

Let {\small $u = Microsoft.ApiCenter/services/workspaces/analyzerConfig/analysisExecutions/read$} \newline
and {\small $v = Microsoft.ApiCenter/deletedServices/delete$}

\medskip
\noindent

The depth of $u$ in $T$ is 7, and the depth of $v$ is 4. Their LCA is {\small \texttt{Microsoft.ApiCenter}} which sits at depth 2. Hence, their distance is 2.

\medskip
\noindent

Here is another example:

Let {\small $u = Microsoft.BotService/botServices/channels/providers/Microsoft.Insights/diagnosticSettings/read$} \newline
and {\small $v = Microsoft.BotService/botServices/channels/providers/Microsoft.Insights/logDefinitions/read$}

\medskip
\noindent

The depth of $u$ and $v$ in $T$ is 9 (notice the two dots in $u$ and $v$; each dot delimitate a segment). 

Their LCA is {\small \texttt{Microsoft.BotService/botServices/channels/providers/Microsoft.Insights}} which sits at depth 7. Hence, their distance is 7.

\subsection{Quantificatying over-approximations}

\subsubsection*{Diameter}
For a permission set $P$, define its diameter as:
$$Diam(P) = \min_{u \neq v \in P} d(u,v)$$

This metric quantifies the "spread" of actions across Azure's Resource Providers hierarchy. Large diameter (small distance between pair of actions) indicates tightly scoped permissions, while small diameter (large distance between pairs) suggests over-approximation.

The reason why we define diameters with a $min$ and not a more conventional $max$ is that we use a linear ultrametric. $max$ is suitable for exponential ultrametrics.

\section{Experimental Evaluation}
\label{sec:evaluation}

\subsection{Compiler}
The python compiler we made for the Azure permissions grammar is called Belshazaar\cite{Bel}. When provided with an \texttt{Action} and a list of \texttt{NotActions}, Belshazaar reads a cache containing all Azure actions and expands the \texttt{Action} and the \texttt{NotActions}. It returns the effective permission set of the action.

Here is an example of a run with $A = Microsoft.AAD/*$ and $N = \{Microsoft.AAD/*/read,Microsoft.AAD/*/delete\}$

\begin{Verbatim}[fontsize=\small]
belshazaar.py --action 'Microsoft.AAD/*' --notActions 'Microsoft.AAD/*/read,Microsoft.AAD/*/delete'

  Microsoft.AAD/domainServices/oucontainer/write
  Microsoft.AAD/domainServices/write
  Microsoft.AAD/register/action
  Microsoft.AAD/domainServices/providers/Microsoft.Insights/diagnosticSettings/write
  Microsoft.AAD/unregister/action
\end{Verbatim}

The effective permission set of $A = Microsoft.AAD/*$ accounting for \texttt{NotActions} is made of 5 explicit permissions, which Belshazaar enumerates.

\subsection{Wildcards Generation}
We wrote a wildcard insertion function which replaces a random subsequence of any action in the official Resource Providers catalog with a wildcard, abiding to the rules of the actions grammar. The exact extent of pattern globbing is identified by the first and last position of the subsequence within the sequence.

Here are a few examples of wildcard generated permissions:

\begin{table}[ht]
\centering
\footnotesize
\begin{tabular}{llll}
\toprule
	Original action & Generated action & first & last \\
\midrule
Microsoft.Blueprint/blueprintAssignments/write & Microsoft.Blueprint/bluepr*s/write& 26 & 38 \\
Microsoft.OperationalInsights/clusters/operationresults/read & Microsoft.Operati*s/read & 17 & 53 \\
Microsoft.Network/networkManagers/routingConfigurations/ruleCollections/rules/delete & Microsoft.Network/networkMana*llections/rules/delete & 28 & 61 \\
\bottomrule
\end{tabular}
	\caption{Examples of wildcard actions generated from original Azure actions}
\label{tab:generation}
\end{table}

\subsection{Extreme Pairs Generation}

To identify highly permissive wildcard patterns in Azure's action space, we developed a wildcard generation pipeline that combines randomized sampling with evolutionary optimization.

\subsubsection{Initial Population}

For each official Azure action string, we first constructed an initial population of 40 candidate wildcard patterns. Wildcards were inserted at random positions within the action string, subject to syntactic constraints designed to avoid trivial global patterns such as \texttt{*} or \texttt{Microsoft.*}.

The following rules were enforced:

\begin{itemize}
    \item The wildcard insertion point must occur at least 3 characters after the Resource Provider dot (the \texttt{.} separator following \texttt{Microsoft}).
    \item Wildcards may not split the final segment of the action unless they fully replace it.
\end{itemize}

This ensures the generation of structurally meaningful wildcards that reflect nontrivial permission generalizations.

\paragraph{Examples:}
\begin{enumerate}
    \item \texttt{Microsoft.Net*/...} is valid because the wildcard is located 4 characters after the dot.
    \item \texttt{Microsoft.O*ions/...} is invalid because the wildcard occurs only 2 characters after the dot.
\end{enumerate}

\subsubsection{Genetic Algorithm Optimization}

After generating the initial population, we applied a Genetic Algorithm (GA) to evolve these wildcard populations over 10 generations. Each population consisted of 40 individuals associated with one original Azure action string.

At each generation, the following steps were performed:

\begin{enumerate}
	\item \textbf{Expansion and Fitness Evaluation:} Each wildcard candidate was expanded using Belshazaar's grammar-based expansion engine, producing the full set of concrete Azure actions it matches. The \emph{Ultrametric Diameter} of each expansion was computed using the ultrametric distance. The fitness function $f(D,x)=1000 \cdot D -x$  was made such as to capture two features in a same scalar: the minimum diameter $D$ and the rightmost wildcard position $x$. Smaller diameters indicate broader wildcard reach (i.e., expansions containing semantically more distant actions), rightmost wildcard positions indicate smaller infix matches.
    
    \item \textbf{Selection:} The population was ranked by $f$ (ascending), and the top 50\% of candidates were selected as survivors for reproduction.
    
    \item \textbf{Mutation and Reproduction:} New candidates were produced by randomly mutating either the start or end position of the wildcard insertion interval. Mutations involved shifting positions by a random positive or negative offset. Each mutated candidate was validated against the original syntactic constraints before inclusion in the population.
\end{enumerate}

This iterative process allowed the algorithm to explore the search space of wildcard placements while avoiding trivial or degenerate cases, and to give a lower bound estimate of the number of extreme pairs.

Our evaluation covered all 15,481 Azure actions retrieved using {\small \texttt{az provider operation list}} on 06 June, 2025.

The full evolutionary search was executed across all official Azure actions, requiring approximately 3.5 hours of computation on a standard desktop machine. The output consists of, for each Azure action, one or more wildcard patterns whose expansions exhibit the smallest ultrametric diameters. These extreme wildcard patterns represent dangerous generalizations in terms of permission breadth.

\subsubsection{Distribution of Diameters}

\paragraph{Median Diameter Computation}
Because ultrametric diameters take on discrete integer values, we report the median via linear interpolation on the empirical cumulative distribution. After sorting the diameter counts and computing the cumulative percentages, we locate the two adjacent diameter values that straddle the 50\% threshold and interpolate between them. This yields a more precise median of 0.98 rather than an integer.

\begin{figure}[htbp]
    \centering
    \includegraphics[width=0.45\textwidth]{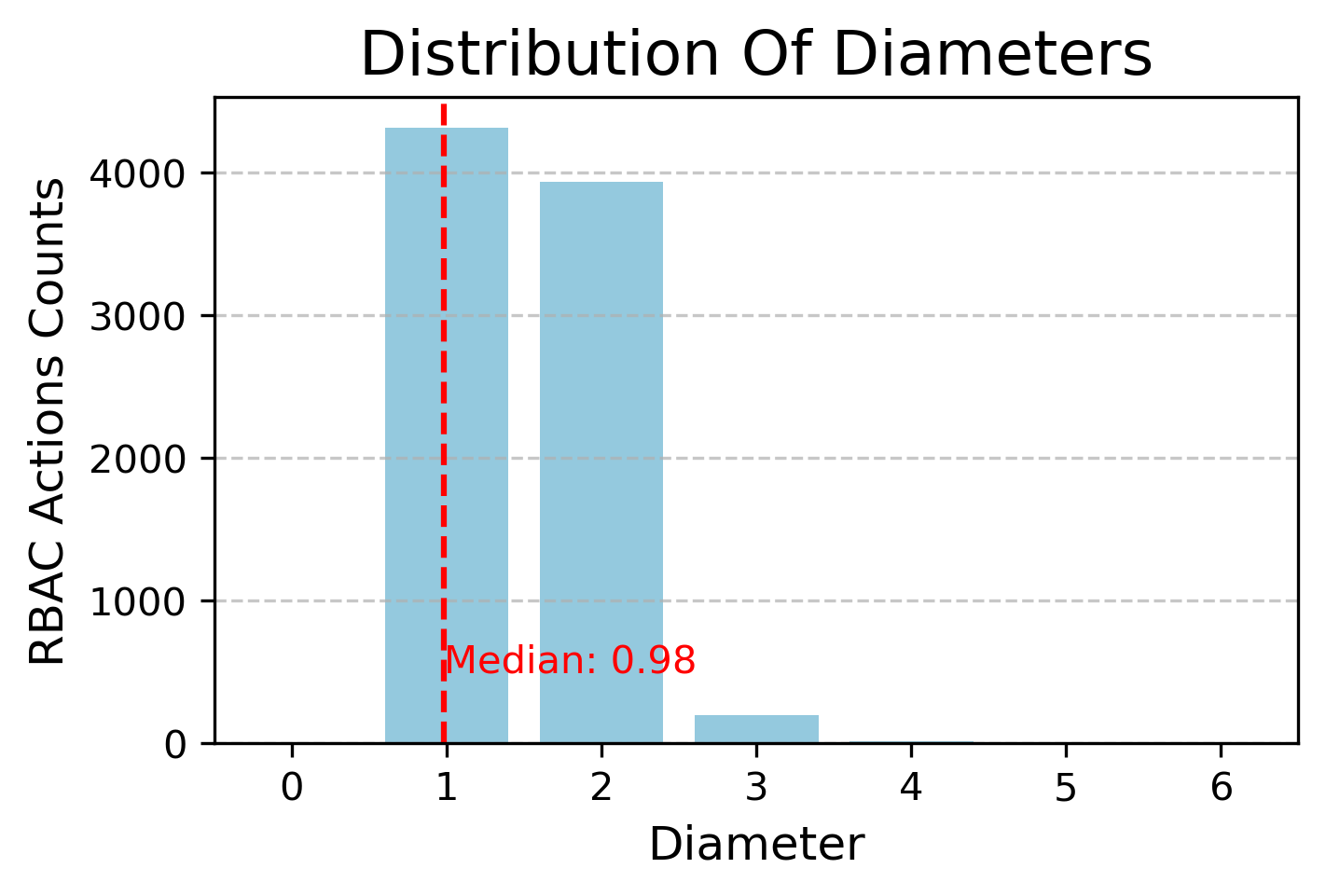}
    \caption{Distribution of Diameters}
    \label{fig:diam}
\end{figure}

This median value of diameters (Figure~\ref{fig:diam}) is extremely low. It is surprising since we took provisions to avoid common obvious wildcards. Recall that a diameter of 1 is indicative of significant over-permissioning wildcard configurations.

\paragraph{Distribution of action pairs by diameter}
Almost all Azure actions can be paired by non-obvious wildcards: about half of all pairs are cross Resource Provider (extreme pairs of diameter 1), another half are same provider, but without least privilege (diameter 2). The number of least privileged pairs (of diameter 3 and above) is less than 3\%. No pairs of diameter greater than 5 were found.

Remember that the number of extreme pairs is a lower bound estimate.

\subsubsection{Wide-ranging wildcards}
Tables ~\ref{tab:toprisks} and ~\ref{tab:lessrisks} provide a small excerpt of extreme pairs generated by wildcards expansion. Table~\ref{tab:toprisks} shows the most extreme pairs, featuring a diameter of 1. Table~\ref{tab:lessrisks} shows pairs featuring diameter 2.

\begin{table}[ht]
\centering
\scriptsize
\begin{tabular}{lll}
\toprule
	Wildcard permission & Left pair & Right pair \\
\midrule
	Microsoft.Api*/write & Microsoft.ApiCenter/services/apis/versions/securityRequirements/write & Microsoft.ApiManagement/gateways/configConnections/write \\
	Microsoft.Aut*s/delete & Microsoft.Authorization/classicAdministrators/delete & Microsoft.Automation/automationAccounts/certificates/delete \\
	Microsoft.Azur*er/action & Microsoft.AzureActiveDirectory/register/action & Microsoft.AzureArcData/register/action \\
	Microsoft.Azure*n/action & Microsoft.AzureDataTransfer/pipelines/approveConnection/action & Microsoft.AzureLargeInstance/AzureLargeInstances/shutdown/action \\
        Microsoft.Cer*ister/action & Microsoft.CertificateRegistration/register/action & Microsoft.Certify/register/action \\
        Microsoft.Contai*start/action & Microsoft.ContainerInstance/containerGroups/restart/action & Microsoft.ContainerService/fleets/updateRuns/start/action \\
        Microsoft.Containe*s/delete & Microsoft.ContainerInstance/containerGroupProfiles/delete & Microsoft.ContainerRegistry/registries/agentpools/delete \\
	Microsoft.Dev*finitions/read & Microsoft.DevCenter/devcenters/catalogs/environmentDefinitions/read & Microsoft.Devices/IotHubs/logDefinitions/read \\
        Microsoft.Data*es/write & Microsoft.DataBoxEdge/dataBoxEdgeDevices/bandwidthSchedules/write & Microsoft.DataFactory/datafactories/datapipelines/write \\
	Microsoft.Hardwar*ers/delete & Microsoft.Hardware/orders/delete & Microsoft.HardwareSecurityModules/cloudHsmClusters/delete \\
        Microsoft.Kubern*s/write & Microsoft.Kubernetes/locations/operationstatuses/write & Microsoft.KubernetesConfiguration/extensions/write \\
        Microsoft.IoT*/write & Microsoft.IoTCentral/IoTApps/privateEndpointConnectionProxies/write & Microsoft.IoTFirmwareDefense/firmwareGroups/firmwares/write \\
        Microsoft.Man*oups/write & Microsoft.ManagedNetworkFabric/neighborGroups/write & Microsoft.Management/managementGroups/write \\
	Microsoft.Net*ps/delete  & Microsoft.NetApp/netAppAccounts/accountBackups/delete & Microsoft.Network/adminNetworkSecurityGroups/delete \\
	Microsoft.Operat*s/read  & Microsoft.OperationalInsights/clusters/operationresults/read & Microsoft.OperationsManagement/managementassociations/read \\
	Microsoft.Netw*nces/delete  & Microsoft.Network/networkSecurityPerimeters/linkReferences/delete & Microsoft.NetworkCloud/storageAppliances/delete \\
\bottomrule
\end{tabular}
	\caption{A sample of extreme (diameter 1) pairs produced by our generator}
\label{tab:toprisks}
\end{table}

\begin{table}[ht]
\centering
\scriptsize
\begin{tabular}{lll}
\toprule
        Wildcard permission & Left pair & Right pair \\
\midrule
	Microsoft.AAD/*tions/read & Microsoft.AAD/Operations/read & Microsoft.AAD/domainServices/providers/Microsoft/Insights/logDefinitions/read \\
	Microsoft.AVS/*/action & Microsoft.AVS/privateClouds/listAdminCredentials/action & Microsoft/AVS/register/action \\
	Microsoft.ApiCenter*s/delete & Microsoft.ApiCenter/deletedServices/delete & Microsoft.ApiCenter/services \\
				     & &  /apis/versions/securityRequirements/delete \\
	Microsoft.Authorization/policy*s/write & Microsoft.Authorization/policyAssignments & Microsoft.Authorization/policyDefinitions/versions/write \\
				     &  /privateLinkAssociations/write &  \\
	Microsoft.Blueprint/bl*/write & Microsoft.Blueprint/blueprintAssignments/write & Microsoft.Blueprint/blueprints/artifacts/write \\
	Microsoft.Cdn/*cies/delete & Microsoft.Cdn/cdnwebapplicationfirewallpolicies/delete & Microsoft.Cdn/profiles/securitypolicies/delete \\
	Microsoft.Cog*s/write & Microsoft.CognitiveServices/accounts/capabilityHosts/write & Microsoft.CognitiveServices/attestations/write \\
	Microsoft.Comp*s/action & Microsoft.Compute/disks/beginGetAccess/action & Microsoft.ComputeSchedule/autoActions/attachResources/action \\
	Microsoft.ContainerSer*snapshots/write & Microsoft.ContainerService/managedclustersnapshots/write  & Microsoft.ContainerService/snapshots/write  \\
	Microsoft.DBforPostgr*n/action & Microsoft.DBforPostgreSQL/assessForMigration/action & Microsoft/DBforPostgreSQL/ \\
	& & flexibleServers/tuningOptions/startSession/action \\
\bottomrule
\end{tabular}
	\caption{A sample of non-extreme (diameter 2) pairs produced by our generator}
\label{tab:lessrisks}
\end{table}

The full list is available on github\cite{azureDiameters}.

\newpage

\section{Limitations and Future Work}

\subsection{Current Limitations}
\begin{itemize}
\item \textbf{Scope}: Limited to Azure RBAC; it does not analyze conditional access or resource-level permissions
\item \textbf{Semantics}: Belshazaar focuses on syntactic analysis; it does not model operational security impact
\end{itemize}

\subsection{Future Directions}
\begin{itemize}
\item \textbf{Multi-Cloud}: Extend framework to AWS IAM and Google Cloud IAM
\item \textbf{Policy Synthesis}: Automatically generate least-privilege roles from usage data
\end{itemize}

\newpage

\section{Conclusion}

In this paper, we presented \textbf{Belshazaar}, a semantics–aware framework for rigorously expanding and evaluating Azure RBAC wildcards. By formalizing the complete Azure action language as a context-free grammar and building a compiler to derive minimal explicit permission sets, Belshazaar delivers an accurate “effective permissions” view for any wildcard specification. Our linear ultrametric diameter model quantifies the semantic overreach of non-obvious wildcards, showing that approximately 50\% span more than one Resource Provider.  

We empirically validated Belshazaar on Microsoft’s catalog of 15,481 actions, demonstrating that it scales to real-world workloads and yields actionable insights into privilege bloat.  

Looking forward, integrating Belshazaar into policy-authoring and continuous-monitoring pipelines will enable administrators to detect and remediate over-privileged roles before deployment. We may extend our ultrametric analysis with user-guided refinement strategies that automatically suggest least-privilege replacements for low-diameter patterns. We believe Belshazaar paves the way toward bridging the gap between policy expressiveness and security assurance in large-scale cloud environments.

\newpage

\section*{Appendix}
\subsection*{Production rules}

Belshazaar's production rules build a normalized Python regex.

\begin{Verbatim}[fontsize=\small]
pattern : segment_list
        p[0] = p[1]

segment_list : segment
        p[0] = p[1]

segment_list : segment_list SLASH segment
        p[0] = p[1] + "/" + p[3]

segment : TEXT
        p[0] = re.escape(p[1])

        | WILDCARD
        p[0] = ".*"

        | TEXT WILDCARD
        p[0] = re.escape(p[1]) + ".*"

        | WILDCARD TEXT
        p[0] = ".*" + re.escape(p[2])

        | TEXT WILDCARD TEXT
	p[0] = re.escape(p[1]) + ".*" + re.escape(p[3])

\end{Verbatim}


\begin{thebibliography}{00}
\bibitem{saltzer1975protection} J. H. Saltzer and M. D. Schroeder, "The protection of information in computer systems," \textit{Proceedings of IEEE}, vol. 63, no. 9, pp. 1278-1308, 1975.

\bibitem{sandhu1996role} R. S. Sandhu, E. J. Coyne, H. L. Feinstein, and C. E. Youman, "Role-based access control models," \textit{Computer}, vol. 29, no. 2, pp. 38-47, 1996.

\bibitem{li2003critique} N. Li, J. C. Mitchell, and W. H. Winsborough, "Design of a role-based trust-management framework," \textit{Proceedings IEEE Symposium on Security and Privacy}, pp. 114-130, 2002.

\bibitem{fisler2005verification} K. Fisler, S. Krishnamurthi, L. A. Meyerovich, and M. C. Tschantz, "Verification and change-impact analysis of access-control policies," \textit{Proceedings International Conference on Software Engineering}, pp. 196-205, 2005.

\bibitem{jackson2012software} D. Jackson, \textit{Software Abstractions: Logic, Language, and Analysis}. MIT Press, 2012.

\bibitem{regular_languages} J. Hopcroft, R. Motwani, and J. Ullman, \textit{Introduction to Automata Theory, Languages, and Computation}, 3rd ed. Pearson, 2006.

\bibitem{azureDiameters} Christophe Parisel, A dump of all Azure action diameters as of June 06, 2025
	\url{https://github.com/labyrinthinesecurity/silhouette/blob/2.1/formal/azureDiameters.txt}


\bibitem{Blast} Christophe Parisel, Scoring Azure permissions with metric spaces \url{https://arxiv.org/abs/2504.13747}

\bibitem{PLY} PLY (PYthon Lex-Yacc), \url{https://ply.readthedocs.io/en/latest/}

\bibitem{Zelkova} How AWS uses automated reasoning to help you achieve security at scale, \url{https://aws.amazon.com/blogs/security/protect-sensitive-data-in-the-cloud-with-automated-reasoning-zelkova/}

\bibitem{Cedar} Cedar, a policy language and evaluation engine, \url{https://github.com/cedar-policy}

\bibitem{Lens} David Kerber, IAM Lens, \url{https://github.com/cloud-copilot/iam-lens}

\bibitem{Bel} Christophe Parisel, Belshazaar, an Azure RBAC actions compiler, \url{https://github.com/labyrinthinesecurity/silhouette/blob/2.1/formal/README.md}

\end{thebibliography}
\end{document}